\documentclass[useAMS,usenatbib,usegraphicx]{mn2e}

\title[A Bayesian Kepler Periodogram Detects a Second Planet in HD 208487]{A Bayesian Kepler Periodogram Detects a Second Planet in HD 208487}
\author[P. C. Gregory]{P. C. Gregory$^{1}$\thanks{E-mail:
gregory@phas.ubc.ca}\footnotemark[1]\thanks{http://www.physics.ubc.ca/~gregory/gregory.html}\\
$^{1}$Physics and Astronomy Department, University of British Columbia,\\ 6224 Agricultural Rd., Vancouver, British Columbia, V6T 1Z1, Canada}
\begin{document}

\date{Submitted 17 Aug. 2006, revised 26 Oct. 2006}

\pagerange{\pageref{firstpage}--\pageref{lastpage}} \pubyear{2006}

\maketitle

\label{firstpage}

\begin{abstract}
An automatic Bayesian Kepler periodogram has been developed for identifying and 
characterizing multiple planetary orbits in precision radial velocity data. The periodogram is powered by a parallel tempering MCMC algorithm which is capable of efficiently exploring a multi-planet model parameter space. The periodogram employs an alternative method for converting the time of an observation
to true anomaly that enables it to handle much larger data sets without a significant 
increase in computation time. Improvements in the periodogram and further tests using data from HD 208487 have resulted in the detection of a second planet with a period of $909_{-92}^{82}$d, an eccentricity of $0.37_{-0.20}^{0.26}$, a semi-major axis of $1.87_{-0.14}^{0.13}$ AU and an $M \sin i = 0.45_{-0.11}^{0.13}$ M$_{\rm J}$. The revised parameters of the first planet are period $= 129.8\pm 0.4$d, eccentricity $= 0.20\pm 0.09$, semi-major axis $= 0.51\pm0.02$ AU and $M \sin i = 0.41\pm0.05$ M$_{\rm J}$.
Particular attention is paid to several methods for calculating the model marginal likelihood which is used to compare the probabilities of 
models with different numbers of planets. 
\end{abstract}


\section{Introduction}

The discovery of multiple planets orbiting the Pulsar PSR B1257+12 (Wolszczan \& Frail, 1992), ushered in an exciting new era of astronomy. Fifteen years later, over 200 extra-solar planets have been discovered by a variety of techniques, including precision radial velocity measurements which have detected the majority of planets to date (Extrasolar Planets Encyclopedia, http://vo.obspm.fr/exoplanetes/encyclo/index.php). It is to be expected that continued monitoring and increased precision will permit the detection of lower amplitude planetary signatures. The increase in parameters needed to model multiple planetary systems is motivating efforts to improve the statistical tools for analyzing radial velocity data (e.g. \citealt{FordGregory2006}, Ford 2005 \& 2006, \citealt{Gregory2005b}, \citealt{Cumming2004}, \citealt{LoredoChernoff2003}, \citealt{Loredo2004}). Much of the recent work has highlighted a Bayesian MCMC approach as a way to better understand parameter uncertainties and degeneracies. 

Gregory (2005a, b \& c) presented a Bayesian MCMC algorithm that makes use of parallel tempering to efficiently explore the full range of model parameter space starting from a random location. It is able to identify any significant periodic signal component in the data that satisfies Kepler's laws and thus functions as a Kepler periodogram~\footnote{Following on from Bretthorst's pioneering work \citep{Brett1988}, many other Bayesian periodograms have been developed. Several examples, based on different prior information, are given by Bretthorst (2001, 2003), \cite{GregoryLoredo1992}, and \citet{Gregory1999}.}. This eliminates the need for a separate periodogram search for trial orbital periods which typically assume a sinusoidal model for the signal that is only correct for a circular orbit. In addition, the Bayesian MCMC algorithm provides full marginal parameters distributions for all the orbital elements that can be determined from radial velocity data. The samples from the parallel chains can also be used to compute the marginal likelihood for a given model \citep{Gregorybook} for use in computing the Bayes factor that is needed to compare models with different numbers of planets. The parallel tempering MCMC algorithm employed in this work includes a control system that automates the selection of efficient Gaussian parameter proposal distributions. This feature makes it practical to carry out blind searches for multiple planets simultaneously.

This paper outlines improvements to the parallel tempering MCMC algorithm that allow for improved mixing and more efficient convergence. In addition several different methods for computing the model marginal likelihood are compared. We also confirm our earlier discovery \citep{Gregory2005MaxEnt} of a second planet in HD 208487.

Some of the analysis presented in this paper was based on the original 31 radial velocity measurements (old data) given in \citet{Tinney2005}. Recently, \citet{Butler2006} published a revised radial velocity data set based on an improved pipeline used to convert the raw spectra to radial velocities. This new data set  also includes 4 new measurements. Most of the results pertain to the new data set, but some of the results dealing with model selection in Section~\ref{sec:modsel} make use of the old data set and this is indicated in the text. 

\section{Parallel tempering}

A simple Metropolis-Hastings MCMC algorithm can run into difficulties if the target probability distribution is multi-modal with widely separated peaks. It can fail to fully explore all peaks which contain significant probability, especially if some of the peaks are very narrow. This is frequently the case with extrasolar planet data which is typically very sparsely sampled. 
The problem is somewhat similar to the one encountered in finding a global $\chi^2$ minimum in a nonlinear model fitting problem. One solution to finding a global minimum is to use simulated annealing by introducing a temperature parameter ${\cal T}$ which is gradually decreased. 

In parallel tempering, multiple copies of the MCMC simulation are run in parallel, each at a different temperature. 
Mathematically, we can describe these tempering distributions by 
\begin{eqnarray}
\pi(X|D,\beta,M_1,I) & = & C\ p(X|M_1,I)p(D|M_1,X,I)^{\beta}\nonumber\\ 
& = & C\ p(X|M_1,I)  \times \nonumber\\
& & \exp(\beta \ \ln[p(D|M_1,X,I)]),\\
& & \mbox{ for } 0 < \beta < 1 \nonumber
\label{eq:lambdaFlatness}
\end{eqnarray}
where $X$ stands for the set of model parameters. The normalization constant, $C$, is unimportant and will be dropped. Rather than use a temperature which varies from 1 to infinity, we use its reciprocal, $\beta=1/{\cal T}$, and refer to as the tempering parameter. Thus $\beta$ varies from 1 to zero.

One of the simulations, corresponding to $\beta  = 1$, is the desired target probability distribution. The other simulations correspond to a ladder of higher temperature distributions. Let $n_\beta$ equal the number of parallel MCMC simulations. At intervals, a pair of adjacent simulations on this ladder are chosen at random and a proposal made to swap their parameter states. A Monte Carlo acceptance rule determines the probability for the proposed swap to occur.  For $\beta=1$, the distribution is the desired target distribution which is referred to as the cold sampler. For $\beta \ll 1$, the distribution is much flatter. This swap allows for an exchange of information across the population of parallel simulations. In the higher temperature simulations, radically different configurations can arise, whereas in higher $\beta$ (lower temperature) states, a configuration is given the chance to refine itself. 
Some experimentation is needed to refine suitable choices of $\beta$ values, which can be assessed by examining the swap acceptance rate between adjacent simulations. If adjacent simulations do not have some overlap the swap rate between them will be very low. The number of $\beta$ values required depends on the application. For parameter estimation purposes a typical value of $n_\beta = 12$. It is important that the iterations from the lowest value of $\beta$ explore the full range of the prior parameter space. For HD 208487, the set $\beta = \{0.05, 0.1, 0.15, 0.25, 0.35, 0.45, 0.55, 0.65, 0.70, 0.80, 0.90, 1.0\}$ proved useful for parameter estimation and achieved a typical acceptance rate between adjacent levels of $\ge 50\%$. The mean number of iterations between swap proposals was set $= 8$. Final inference is based on samples drawn from the $\beta = 1.0$ simulation. 

It is possible to use the samples from hotter simulations to evaluate the marginal (global) likelihood needed for model selection (see Section 12.7 of \citealt{Gregorybook}). Marginal likelihoods estimated in this way typically require many more parallel simulations. For HD 208487, 34 $\beta$ levels were used spanning the range $\beta = 10^{-8}$ to $1.0$. This is discussed more in Section~\ref{sec:modsel}.

\subsection{Proposal distributions}

In Metropolis-Hastings versions of MCMC, parameter proposals are drawn from a proposal distribution. In this analysis, independent Gaussian proposal distributions, one for each parameter, were employed. Of course, for sparse data sets it can often be the case that some of the parameters are highly correlated resulting in inefficient sampling from independent Gaussians. One solution is to use combinations of model parameters that are more independent. More on this later. In general, the $\sigma$'s of these Gaussian proposal distributions are different because the parameters can be very different entities. Also if the $\sigma$'s are chosen to small, successive samples will be highly correlated and it will require many iterations to obtain an equilibrium set of samples. If the $\sigma$'s are too large, then proposed samples will very rarely be accepted. Based on empirical studies, \citet{Roberts1997} recommend calibrating the \index{MCMC!acceptance rate}acceptance rate to about $25\%$ for a high-dimensional model and to about $50\%$ for models of 1 or 2 dimensions. Although these studies were based on a multinormal target probability distributions, they have proven a useful guideline for our application as well.  

The process of choosing a set of useful proposal $\sigma$'s when dealing with a large number of different parameters can be very time consuming. In  parallel tempering MCMC, the problem is compounded because of the need for a separate set of proposal $\sigma$'s for each simulation. We have automated this process using a two stage statistical control system (CS) in which the error signal is proportional to the difference between the current acceptance rate and a target acceptance rate, typically 25\%. In the first stage an initial set of proposal $\sigma$'s ($\approx 10\%$ of the prior range for each parameter) are separately perturbed to determine an approximate gradient in the acceptance rate with respect to the proposal $\sigma$'s. The $\sigma$'s are then jointly modified by a small increment in the direction of this gradient. This is done for each of the parallel simulations or chains as they are sometimes called. When the $\sigma$'s are large all the MCMC simulations explore the full prior distribution and locate significant probability peaks in the joint parameter space. As the proposal $\sigma$'s decrease these peaks are more efficiently explored in the $\beta = 1$ simulation. This annealing of the proposal $\sigma$'s typically takes place over the first 5,000 to 150,000 (unthinned) iterations for one planet and first 5,000 to 300,000 iterations for two planets. This may seem like an excessive number of iterations but keep in mind that we are dealing with sparse data sets that can have multiple, widely separated probability peaks and we want the MCMC to locate the most significant probability peak before finalizing the choice of proposal $\sigma$'s. 

Although the acceptance rate for the final joint set of parameter $\sigma$'s is achieved, it often happens that a subset of the proposal $\sigma$'s will be too small leading to excessive correlation in the MCMC iterations for these parameters. The second stage CS corrects for this. 
In general, the burn-in period occurs within the span of the first stage CS, i.e., the significant peaks in the joint parameter probability distribution are found, and the second stage improves the choice of proposal $\sigma$'s for the highest probability parameter set. Occasionally, a new higher probability parameter set emerges at a later iteration. The second phase of the control system can detect this and compute a new set of proposal $\sigma$'s. If this happens the control system resets the burn-in period to include all previous iterations.
The useful MCMC simulation data is obtained after the CS is switched off.
Although inclusion of the control system may result in a somewhat longer effective burn-in period, there is a huge saving in time because it eliminates many trial runs to manually establish a suitable set of proposal $\sigma$'s.

\section{Re-parameterization}

Ford (2006) examined the effect of a variety of re-parameterizations for the Kepler model on the MCMC convergence speed. By far the biggest improvement was obtained with re-parameterizations that involved $\omega$, the argument of periastron, and $M_0$, the mean anomaly at initial epoch. This is because for low eccentricity orbits $\omega$ can be poorly constrained but the observational data can often better constrain the $\omega+M_0$.

In our analysis the predicted radial velocity is given 
\begin{equation}
v(t_i) = V + K [\cos\{\theta(t_i+\chi P)+\omega\} + e \cos \omega],
\label{eq:orbit1}
\end{equation}
and involves the 6 unknowns
\begin{itemize}
\item[] $V =$ a constant velocity.
\item[] $K =$ velocity semi-amplitude. 
\item[] $P =$ the orbital period.
\item[] $e =$ the orbital eccentricity.
\item[] $\omega =$ the longitude of periastron.
\item[] $\chi =$ the fraction of an orbit, prior to the start of data taking, that periastron occurred at. Thus, $\chi P =$ the number of days prior to $t_i = 0$ that the star was at periastron, for an orbital period of P days. 
\item[] $\theta(t_i+\chi P) =$ the angle of the star in its orbit relative to periastron at time $t_i$, also called the true anomaly.
\end{itemize}

We utilize this form of the equation because we obtain the dependence of $\theta$ on $t_i$ by solving the conservation of angular momentum equation
\begin{equation}
\frac{d\theta}{dt} - \frac{2 \pi [1+e\cos \theta(t_i+\chi \; P)]^2}{P (1-e^2)^{3/2}} = 0.
\label{eq:orbit2}
\end{equation}
Our algorithm is implemented in {\it Mathematica} and it proves faster for {\it Mathematica} to solve this differential equation than solve the equations relating the true anomaly to the mean anomaly via the eccentric anomaly. {\it Mathematica} generates an accurate interpolating function between $t$ and $\theta$ so the differential equation does not need to be solved separately for each $t_i$. Evaluating the interpolating function for each $t_i$ is very fast compared to solving the differential equation, so the algorithm should be able to handle much larger samples of radial velocity data than those currently available without a significant increase in computational time.

Instead of varying $\chi$ and $\omega$ at each MCMC iteration, we varied $\psi=2\pi\chi+\omega$ and $\phi=2 \pi\chi-\omega$, motivated by the work of Ford (2006).
$\psi$ is well determined for all eccentricities. Although $\phi$ is not well determined for low eccentricities, it is at least orthogonal to the $\psi$ parameter. It is easy to demonstrate that uniform sampling of $\psi$ in the interval 0 to $2 \pi$ and uniform sampling of $\phi$ in the interval $-2 \pi$ to $+2 \pi$ results in uniform coverage of $\chi$ in the interval 0 to 1 and uniform coverage of $\omega$ from 0 to $2 \pi$ using the relations
\begin{eqnarray}
\chi & = & {\rm Modulus} \left[\frac{\psi+\phi}{4 \pi},1 \right] \nonumber\\
\omega & = & {\rm Modulus} \left[\frac{\psi-\phi}{2},2 \pi \right]
\label{eq:reparm}
\end{eqnarray}
Restricting $\phi$ to the interval 0 to $2 \pi$ produces an hourglass coverage of half the $\chi, \omega$ plane. 
This re-parameterization, together with the additional second stage CS, achieved good mixing of the MCMC iterations for a wide range of orbital eccentricities. 

\section{Frequency search}

For the Kepler model with sparse data, the target probability distribution can be very spiky. This is particularly a problem for the orbital period parameters which span roughly 6 decades.  In general, the sharpness of the peak depends in part on how many periods fit within the duration of the data. The previous implementation of the parallel tempering algorithm employed a proposal distribution for $P$ which was a Gaussian in the logarithmic of $P$. This resulted in a constant fractional period resolution instead of a fixed absolute resolution, increasing the probability of detecting narrow spectral peaks at smaller values of $P$. However, this proved not to be entirely satisfactory because for the HD 73526 data set of \citet{Tinney2003}, one of the three probability peaks (the highest) was not detected in two out of five trials \citep{Gregory2005b}. 

Our latest algorithm implements the search in frequency space for the following reasons.
In a Bayesian analysis, the width of a spectral peak, which reflects the accuracy of the frequency estimate, is determined by the duration of the data, the signal-to-noise (S/N) ratio and the number of data points.  More precisely (\citealt{Gregorybook}, \citealt{Brett1988}), for a sinusoidal signal model, the standard deviation of the spectral peak, $\delta f$, for a S/N $>1$, is given by
\begin{equation}
\delta f \approx  \left(1.6 \frac{{\rm S}}{{\rm N}} {\cal{T}} \sqrt{N}\right)^{-1} \ \ {\rm{Hz}},
\label{eq:Bfres}
\end{equation}
where $\cal{T} =$ the data duration in s, and $N =$ the number of data points in $\cal{T}$. The thing to notice is that the width of any peak is independent of the frequency of the peak. Thus the same frequency proposal distribution will be efficient for all frequency peaks. This is not the case for a period search where the width of a spectral peak is $\propto P^2$. Not only is the width of the peak independent of f,
but the spacing of peaks is constant in frequency (roughly $\Delta f \sim 1/T$), which
is a another motivation for searching in frequency space (e.g., \citealt{Scargle1982}, \citealt{Cumming2004}). With a frequency search strategy, a re-analysis of the original HD 73526 data set resulted in all three peaks being detected in five out of five trials.

\section{Choice of Priors}
\label{sec:priors}

In a Bayesian analysis we need to specify a suitable prior for each parameter. We first address the question of what prior to use for frequency for multi-planet models. In this work, the lower cutoff in period of 1d is chosen to be less than the smallest orbital period of known planets but somewhat larger than the Roche limit which occurs at 0.2d for a $10M_{Jup}$ planet around a $1M_{\sun}$ star. The upper period  period cutoff of 1000yr is much longer than any known extrasolar planet, but corresponds roughly to a period where perturbations from passing stars and the galactic tide would disrupt the planet's orbit \citep{FordGregory2006}. For a single planet model we use a Jeffreys prior because the prior period (frequency) range spans almost 6 decades. A Jeffreys prior corresponds to a uniform probability density in $\ln f$. This says that the true frequency is just as likely to be in the bottom decade as the top. The Jeffreys prior can be written in two equivalent ways.
\begin{equation}
p(\ln f|M_1,I)\ d\ln f = \frac{d\ln f}{\ln (f_H/f_L)}
\label{eq:Jeff1}
\end{equation}
\begin{equation}
p(f|M,I)\ df = \frac{df}{f\  \ln (f_H/f_L)}
\label{eq:Jeff2}
\end{equation}

What form of frequency prior should we use for a multiple planet model? We first develop the prior to be used in a frequency search strategy where we constrain the frequencies in an $n$ planet search such that $(f_L \le f_1 \le f_2 \cdots \le f_n \le f_H)$. From the product rule of probability theory and the above frequency constraints we can write
\begin{eqnarray}
& & p(\ln f_1, \ln f_2, \cdots \ln f_n|M_n,I) = p(\ln f_n|M_n,I)\nonumber\\
& & \ \ \ \ \ \times p(\ln f_{n-1}|\ln f_n,M_n,I)\cdots p(\ln f_2|\ln f_3,M_n,I)\nonumber\\
& & \ \ \ \ \ \times p(\ln f_1|\ln f_2,M_n,I).
\label{eq:freqs}
\end{eqnarray} 
For model selection purpose we need to use a normalized prior which translates to the requirement that
\begin{equation}
\int_{\ln f_L}^{\ln f_H} p(\ln f_1, \ln f_2, \cdots \ln f_n|M_n,I) d\ln f_1 \cdots d\ln f_n = 1.
\label{eq:freqs1}
\end{equation} 
We assume that $p(\ln f_1, \ln f_2, \cdots \ln f_n|M_n,I)$ is equal to a constant $k$ everywhere within the prior volume. We can solve for $k$ from the integral equation
\begin{equation}
k \int_{\ln f_L}^{\ln f_H} d\ln f_n \int_{\ln f_L}^{\ln f_n} d\ln f_{n-1} \cdots \int_{\ln f_L}^{\ln f_2} d\ln f_1  = 1.
\label{eq:freqs2}
\end{equation}
The solution to equation~(\ref{eq:freqs2}) is 
\begin{equation}
k = \frac{n!}{[\ln (f_H/f_L)]^n}.
\label{eq:freqs3}
\end{equation}
The joint frequency prior is then
\begin{equation}
p(\ln f_1, \ln f_2, \cdots \ln f_n|M_n,I) = \frac{n!}{[\ln (f_H/f_L)]^n} 
\label{eq:freqs4}
\end{equation} 
Expressed as a prior on frequency, equation~(\ref{eq:freqs3}) becomes
\begin{equation}
p(f_1,f_2, \cdots f_n|M_n,I) = \frac{n!}{f_1 f_2 \cdots f_n\  [\ln (f_H/f_L)]^n} 
\label{eq:freqs5}
\end{equation} 
We note that a similar result, involving the factor $n!$ in the numerator, was obtained by Bretthorst (2003) in connection with a uniform frequency prior.  

Two different approaches to searching in the frequency parameters were employed in this work. In the first approach (a): an upper bound on $f_1 \le f_2$ ($P_2 \ge P_1$) was utilized to maintain the identity of the two frequencies. In the second more successful approach (b): both $f_1$ and $f_2$ were allowed to roam over the entire frequency range  and the parameters re-labeled afterwards. In this second approach nothing constrains $f_1$ to always be below $f_2$ so that degenerate parameter peaks can occur. For a two planet model there are twice as many peaks in the probability distribution possible compared with (a). For a $n$ planet model, the number of possible peaks is $n!$ more than in (a). Provided the parameters are re-labeled after the MCMC, such that parameters associated with the lower frequency are always identified with planet one and vice versa, the two cases are equivalent~\footnote{To date this claim has been tested for $n \le 3$.} and equation~(\ref{eq:freqs4}) is the appropriate prior for both approaches.

Approach (b) was found to be more successful because in repeated blind period searches it always converged on the highest posterior probability distribution peak, in spite of the huge period search range. Approach (a) proved to be unsuccessful in finding the highest peak in some trials and in those cases where it did find the peak it required many more iterations. Restricting $P_2 \ge P_1$ ($f_1 \le f_2$) introduces an additional hurdle that appears to slow the MCMC period search. 

The full set of priors used in our Bayesian calculations are given in Table~\ref{tab:priors}. 
Two different limits on $K_i$ were employed.  In the first case \#1, the upper limit corresponds to the velocity of a planet with a mass $= 0.01$ M$_{\sun}$ in a circular orbit with our shortest period of one day. Also the upper bound on $P_i$ of 1000 yr is an upper bound based on galactic tidal disruption. Previously we used an upper limit of three times the duration of the data. 
An upper bound of $K_{\rm max}\ \left(\frac{P_{\rm min}}{P_i}\right)^{1/3}$ was proposed at an exoplanet workshop at the Statistics and Applied Math Sciences Institute (spring 2006), however, the factor of $\left(\frac{P_{\rm min}}{P_i}\right)^{1/3}$ was not incorporated in early runs of the current analysis. We set $K_{\rm max} = 2129$m s$^{-1}$, which corresponds to a maximum planet-star mass ratio of 0.01.

For case \#2, the upper limit on $K_i$ was set equal to $K_{\rm max}\ \left(\frac{P_{\rm min}}{P_i}\right)^{1/3} \frac{1}{\sqrt{1-e_i^2}}$ based on equation~(\ref{eq:Kequ}).
\begin{equation}
K=\frac{m \sin i}{M_*}\left(\frac{2 \pi G M_*}{P}\right)^{1/3} \left(1+\frac{m}{M_*}\right)^{-2/3} \frac{1}{\sqrt{1-e_i^2}},
\label{eq:Kequ}
\end{equation}
where $m$ is the planet mass, $M_*$ is the star's mass, and $G$ is the gravitational constant. Case \#2 is an improvement over $K_{\rm max}\ \left(\frac{P_{\rm min}}{P_i}\right)^{1/3}$ because it allows the upper limit on $K$ to depend on the orbital eccentricity. Clearly, the only chance we have of detecting an orbital period of 1000 yr with current data sets is if the eccentricity is close to one and we are lucky enough to capture periastron passage. Prior \#2 was used in this work with the exception of Section~\ref{sec:modsel} on model selection. In that section, some results were obtained with prior \#1 and the rest with prior \#2, as indicated in the text. 
\begin{table*}
 \centering
 \begin{minipage}{140mm}
  \caption{Prior parameter probability distributions.}
  \label{tab:priors}
  \begin{tabular}{@{}llll@{}}
  \hline
   Parameter    &    Prior        & Lower bound & Upper bound\\
 \hline
Orbital period $P_i$  & $\frac{n! \ P^{-1}}{\ln{\left(\frac{P_{\rm max}}{P_{\rm min}}\right)}}$  & 1.0 d & 1000 yr  \\
& & & \\
Velocity $K_i$  & (1) Modified Jeffreys~\footnote{Since the prior lower limits for $K$ and $s$ include zero, we used a modified Jeffreys prior of the form
\begin{equation}
p(X|M,I) = \frac{1}{X+X_0}\; \frac{1}{\ln\left(1+\frac{X_{\rm max}}{X_0}\right)}
\label{eq:orbit13}
\end{equation}
For $X \ll X_0$, $p(X|M,I)$ behaves like a uniform prior and for $X \gg X_0$ it behaves like a Jeffreys prior. The $\ln\left(1+\frac{ X_{\rm max}}{X_0}\right)$ term in the denominator ensures that the prior is normalized in the interval 0 to $X_{\rm max}$.} & 0 \ (K$_0 = 1)$ &  $K_{\rm max}=2129$ \\
\ \ \  (m s$^{-1}$) & & & \\
  & \ \ \ \ \ $\frac{1}{K+K_0}\; \frac{1}{\ln\left(1+\frac{K_{\rm max}}{K_0}\right)}$
 &  & \\
 & & & \\
& (2) $\frac{(K+K_0)^{-1}}{\ln{\left[1+\frac{K_{\rm max}}{K_0} \ \left(\frac{P_{\rm min}}{P_i}\right)^{1/3} \frac{1}{\sqrt{1-e_i^2}}\right]}}$ & 0 & $K_{\rm max}\ \left(\frac{P_{\rm min}}{P_i}\right)^{1/3} \frac{1}{\sqrt{1-e_i^2}}$ \\
& & & \\
V  (m s$^{-1}$) & $-K_{\rm max}$ & $K_{\rm max}$ &   \\
& & & \\
Eccentricity $e_i$ & Uniform & 0 & 1 \\
& & & \\
& & & \\
Longitude of periastron $\omega_i$ & Uniform & $-2 \pi$ & $2 \pi$ \\
& & & \\
Extra noise $s$  (m s$^{-1}$) & $\frac{(s+s_0)^{-1}}{\ln{\left(1+\frac{s_{\rm max}}{s_{0}}\right)}}$ & 0 & $K_{\rm max}$  \\
\ \ standard deviation & \ \ \ \ (s$_0 = 1\ \& \ 10$m s$^{-1}$)& & \\
\hline
\end{tabular}
\end{minipage}
\end{table*}

Three of the models considered in this paper, $M_0$ (no planet), $M_1$ (one planet), and $M_2$ (two planet model), incorporate an extra noise parameter, $s$, that can allow for any additional noise beyond the known measurement uncertainties~\negthinspace\footnote{In the absence of detailed knowledge of the sampling distribution for the extra noise, we pick a Gaussian because, for any given finite noise variance, it is the distribution with the largest uncertainty as measured by the entropy, i.e., the maximum entropy distribution (\citealt{Jaynes1957}, \citealt{Gregorybook} section 8.7.4.)}. We assume the noise variance is finite and adopt a Gaussian distribution with a variance $s^2$. Thus, the combination of the known errors and extra noise has a Gaussian distribution with variance $= \sigma_i^2 + s^2$, where $\sigma_i$ is the standard deviation of the known noise for i$^{\mbox{\tiny th}}$ data point. For example, suppose that the star actually has two planets, and the model assumes only one is present. In regard to the single planet model, the velocity variations induced by the unknown second planet acts like an additional unknown noise term. Other factors like star spots and chromospheric activity can also contribute to this extra velocity noise term which is often referred to as stellar jitter. Several researchers have attempted to estimate stellar jitter for individual stars based on statistical correlations with observables (e.g., \citealt{Saar1997}, \citealt{Saar1998}, \citealt{Wright2005}). In general, nature is more complicated than our model and known noise terms. Marginalizing $s$ has the desirable effect of treating anything in the data that can't be explained by the model and known measurement errors as noise, leading to conservative estimates of orbital parameters (see Sections 9.2.3 and 9.2.4 of \citet{Gregorybook} for a tutorial demonstration of this point). If there is no extra noise then the posterior probability distribution for $s$ will peak at $s = 0$. The upper limit on $s$ was set equal to $K_{\rm max}$. For the old data set we employed a modified Jeffrey's prior with a knee, $s_0 = 1$m s$^{-1}$. For the new data we carried out the calculations for two different choices, namely, $s_0 = 1$ and $s_0 = 10$m s$^{-1}$. 

We also consider a fourth model $M_{1_j}$, a one planet model with a Gaussian prior for the extra noise parameter $s$ of the form
\begin{equation}
p(s|M_{1_j},I) = k_J \exp{\left(-\frac{(s-s_a)^2}{2 \sigma_s^2} \right)},
\label{eq:priorM1J}
\end{equation}
where $k_J$ is the normalization constant given by
\begin{equation}
 k_J = 1/\int_0^{s=K_{\rm max}}\exp{\left(-\frac{(s-s_a)^2}{2 \sigma_s^2} \right)} ds.
\label{eq:priorM1Jnorm}
\end{equation}
For this model we set $s_a = 5.4$m s$^{-1}$, the jitter estimate for HD 208487 given by \citet{Butler2006}, based on Wright's (2005) jitter modeling. We set $\sigma_s = 3$m s$^{-1}$ as an estimate of the uncertainty in the value of $s_a$.

\section{Results}

\begin{figure}
\includegraphics[width=80mm]{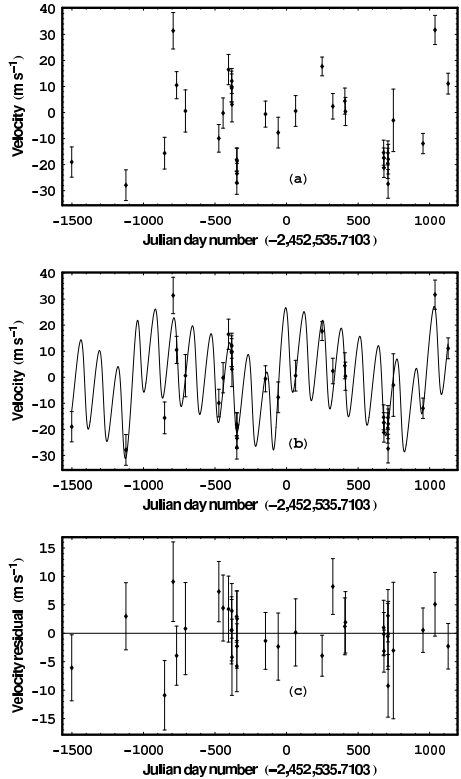}
 \caption{The new data is shown in panel (a) and the best fitting two planet ($P_1 = 129.8$ day, $P_2 = 908$ day) model versus time is shown in (b). Panel (c) shows the residuals.}
\label{fig:dattimres}
\end{figure}

As mentioned in the introduction, the initial analysis was carried out using the 31 radial velocity measurements from Tinney et al. (2005) who reported the detection of a single planet with $M \sin i = 0.45\pm0.05$ in a $130\pm1$ day orbit with an eccentricity of $0.32\pm0.1$. Figure~\ref{fig:dattimres} shows the new precision radial velocity data for HD 208487 from \citet{Butler2006} who reported a single planet with $M \sin i = 0.52\pm0.08$ in a $130.08\pm0.51$ day orbit with an eccentricity of $0.24\pm0.16$. The additional data points are the last 4 shown in the figure. The new pipeline resulted in changes to the values of the original 31 data points. The largest change after allowance for the different means (different zero points) was 6.9 m s$^{-1}$ or 1.3 $\sigma$. 

\citet{Gregory2005MaxEnt} reported results from a preliminary re-analysis of the old data set which indicated a second planet with a period of $\approx 1000$ days. Panels (b) and (c) show the best fitting two planet light curve and residuals based on the more detailed analysis of the new data which is presented in this paper. 

\begin{figure}
\includegraphics[width=80mm]{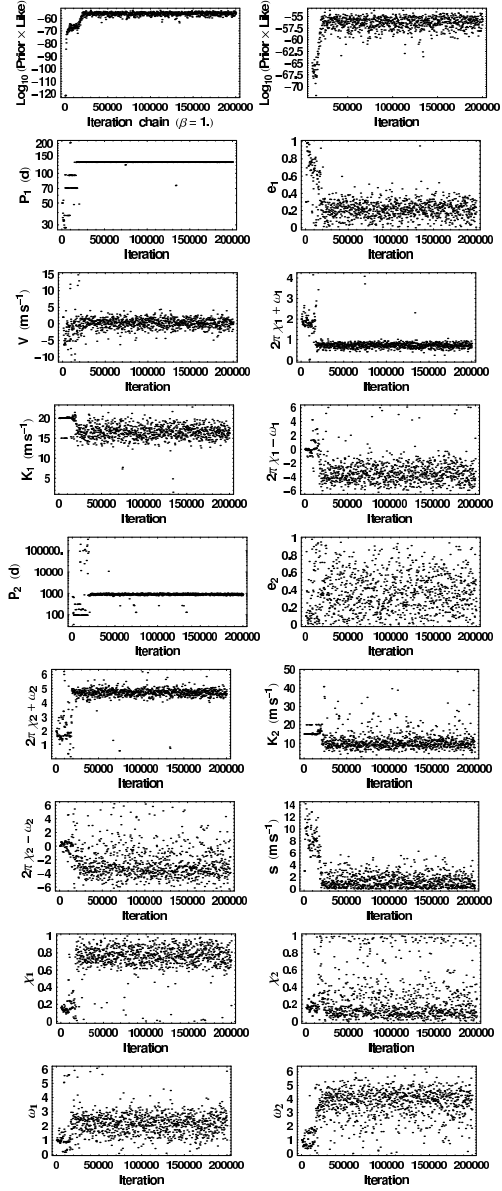}
 \caption{MCMC parameter iterations. The upper left panel is a plot of the prior $\times$ likelihood, and the upper right panel shows a blow-up of the y-axis after dropping the first 10,000 iterations.}
\label{fig:iter}
\end{figure}

\begin{figure}
\includegraphics[width=80mm]{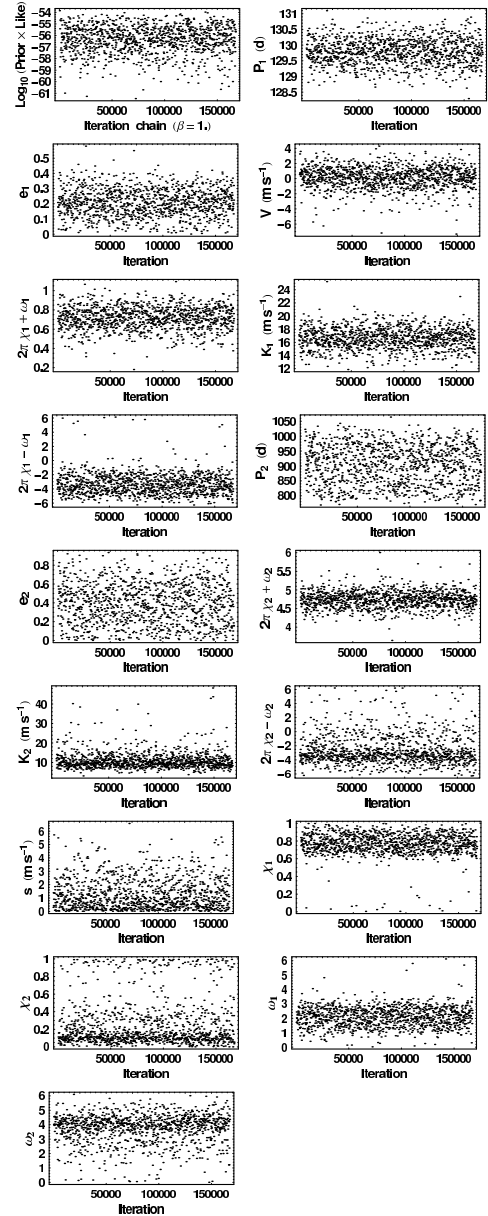}
 \caption{Post burn-in MCMC iterations for a window in period space ($125 \le P_1 \le 135$d and $650\le P_1 \le 1200$d) that isolates the dominant peak.}
\label{fig:iterbest}
\end{figure}

\begin{figure}
\includegraphics[width=80mm]{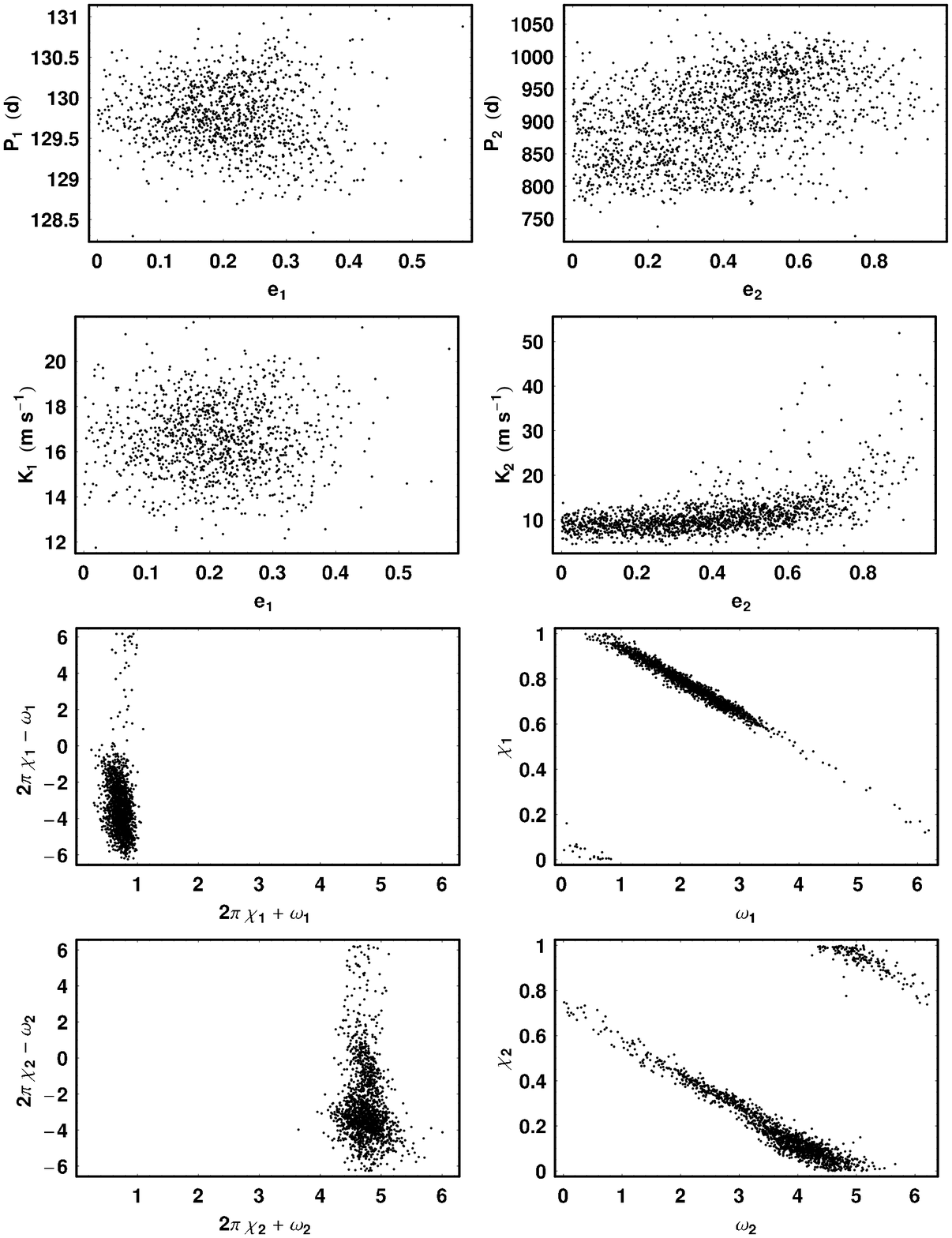}
 \caption{Joint marginals for various pairs of parameter values.}
\label{fig:corr}
\end{figure}

Figure~\ref{fig:iter} shows $\beta =1$ MCMC iterations for each of the parameters starting from a specific but arbitrary initial location in parameter space of $P_1 = 50$ d, $e_1=0.3$, $V=2.0$ ms$^{-1}$, $\psi_1=2.0$ radians, $K_1=20$ ms$^{-1}$, $\phi_1=0.0$ radians, $P_2=700$ d, $e_2=0.1$, $\psi_2=2.0$ radians, $K_2=15$ ms$^{-1}$, $\phi_2=0.0$ radians, $s=3$ ms$^{-1}$. A total of 1.8 million iterations were used but only every sixth iteration was stored. For display purposes only every two hundredth point is plotted in the figure. The burn-in period of approximately 40,000 iterations is clearly discernable. The dominant solution corresponds to $P_1=129.8$d and $P_2=908$d, but there are occasional jumps to other remote periods demonstrating that the parallel tempering algorithm is exploring the full prior range in search of other peaks in the target posterior distribution. The $\chi_i$ and $\omega_i$ traces were derived from the corresponding $\psi_i,\phi_i$ traces using equation~(\ref{eq:reparm}). Figure~\ref{fig:iterbest} shows the post burn-in iterations for a window in period space ($125 \le P_1 \le 135$d and $650\le P_1 \le 1200$d) that isolates the dominant peak. All the traces appear to have achieved an equilibrium distribution. There is a weak correlation between $P_2$ and $e_2$ and weak correlation tail evident between $K_2,e_2$, and $V$. These correlations are shown better in the joint marginal distributions for 6 pairs of parameters in Figure~\ref{fig:corr}. Each dot is the result from one iteration. 
 The lower four panels of this plot nicely illustrate the advantage of the using the $\psi_i = 2 \pi \chi_i + \omega_i$ and $\phi_i = 2 \pi \chi_i - \omega_i$ re-parameterization, which are essentially uncorrelated in comparison to the $\chi_i$ and $\omega_i$.   
 

\begin{figure}
\includegraphics[width=80mm]{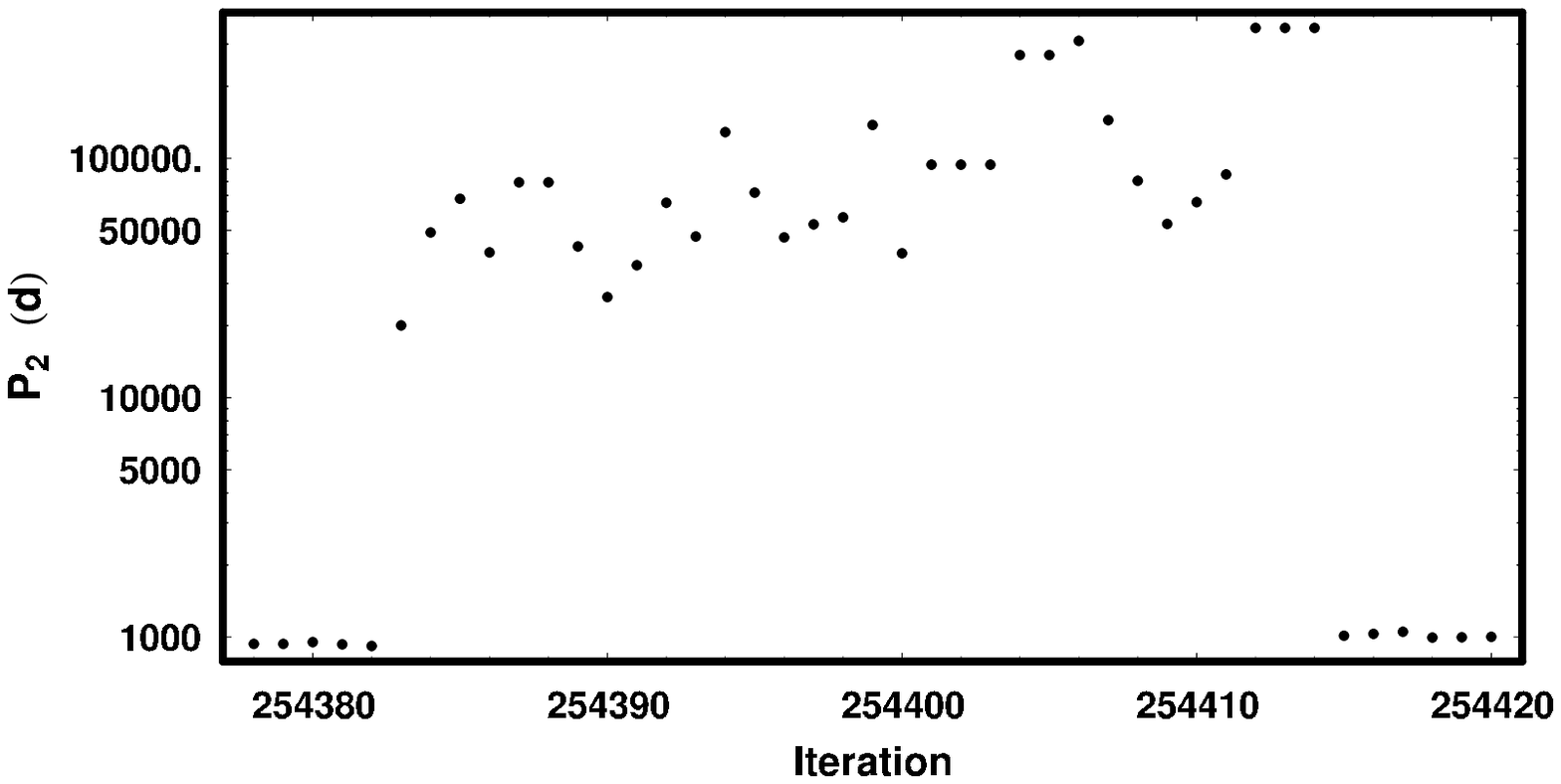}
 \caption{A blow-up of a small range of parameter $P_2$ iterations illustrating a transition to a location that is very remote from the dominant solution that occasionally results from the parallel tempering swap operation. This allows the algorithm to widely explore the full prior parameter space in search of other significant peaks.}
\label{fig:swapjump}
\end{figure}

The Gelmen-Rubin statistic is typically used to test for convergence of the parameter distributions. In parallel tempering MCMC, new widely separated parameter values are passed up the line to the $\beta = 1$ simulation and are occasionally accepted. Roughly every 100 iterations the $\beta = 1$ simulation accepts a swap proposal from its neighboring simulation. If the transition is to a location in parameter space that is very remote from the dominant solution, then the $\beta = 1$ simulation will have to wait for another swap to return it to the dominant peak region. One example of such a swapping operation is shown for the $P_2$ parameter in Figure~\ref{fig:swapjump}. Of course most swaps are to locations within the equilibrium distribution of the dominant peak. 
The final $\beta = 1$ simulation is thus an average of a very large number of independent $\beta = 1$ simulations.  What we have done is divide the $\beta = 1$ iterations into ten equal time intervals and inter compared the ten different essentially independent average distributions for each parameter using a Gelmen-Rubin test. For all of the two planet model parameters the Gelmen-Rubin statistic was $\le 1.02$.
\begin{figure}
\includegraphics[width=80mm]{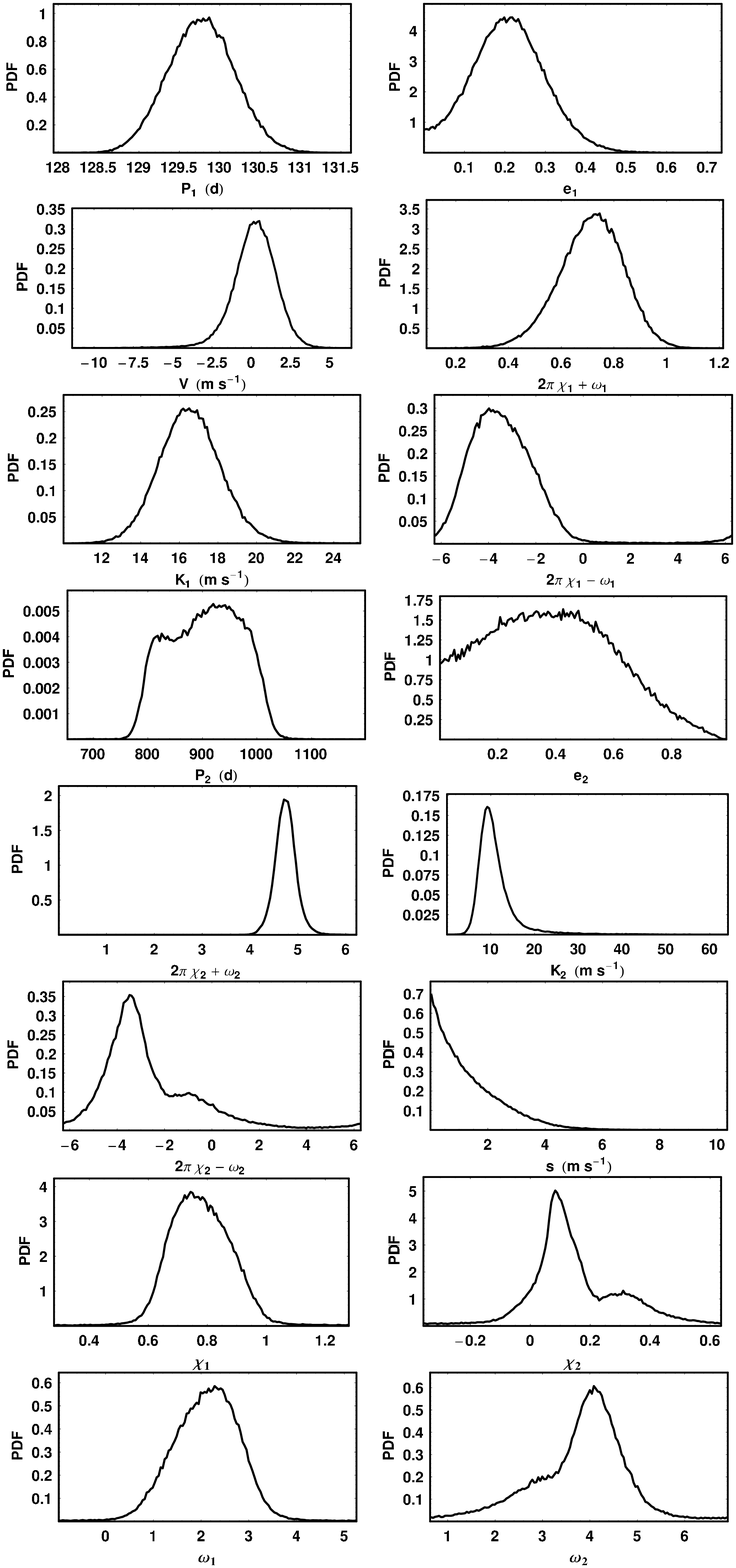}
\caption{Marginal parameter probability distributions for the two planet model for the new data set and $s_0=1$m s$^{-1}$.}
\label{fig:marg2}
\end{figure}
\begin{figure}
\includegraphics[width=80mm]{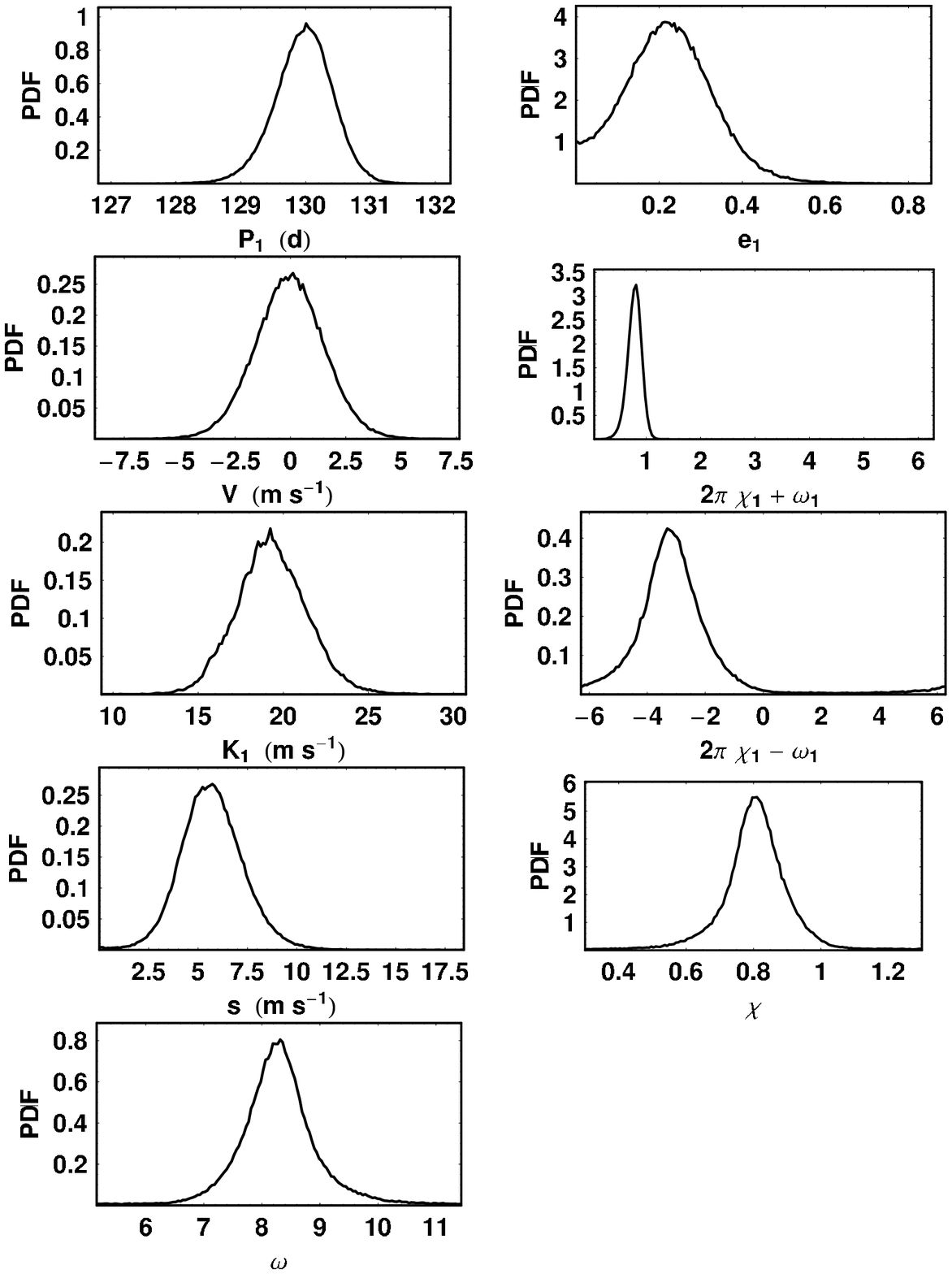}
\caption{Marginal parameter probability distributions for the one planet model for the new data set and $s_0=1$m s$^{-1}$.}
\label{fig:marg1}
\end{figure}

Figure~\ref{fig:marg2} shows the individual parameter marginal distributions for the two planet model dominant solution. For comparison purposes, the marginals distributions for the one planet model are shown in Figure~\ref{fig:marg1}. Table~\ref{tab:parerrorsM1} compares our Bayesian one planet orbital parameter values and their errors, for the two different choices of $s_0$, to the values from a) Tinney et al. (2005) and b) Butler et al. (2006). The parameter values given for our analysis are the median of the marginal probability distribution for the parameter in question and the error bars identify the boundaries of the 68.3\% credible region. The value immediately below in parenthesis is the MAP value, the value at the maximum of the joint posterior probability distribution. It is clear from Table~\ref{tab:parerrorsM1} that changing $s_0$ from 1 to 10m s$^{-1}$ did not significantly alter the $M_1$ parameter estimates. The values derived for the semi-major axis and $M \sin i$, and their errors, are based on the assumed mass of the star $= 1.05\pm0.12$ M$_{\sun}$ \citep{Valenti2005}. \citet{Tinney2005} assumed a mass of $= 0.95\pm0.05$ M$_{\sun}$, while \citet{Butler2006} assumed a mass of $= 1.13$ M$_{\sun}$ but also quote \citet{Valenti2005} as the reference. 

\begin{figure}
\includegraphics[width=80mm]{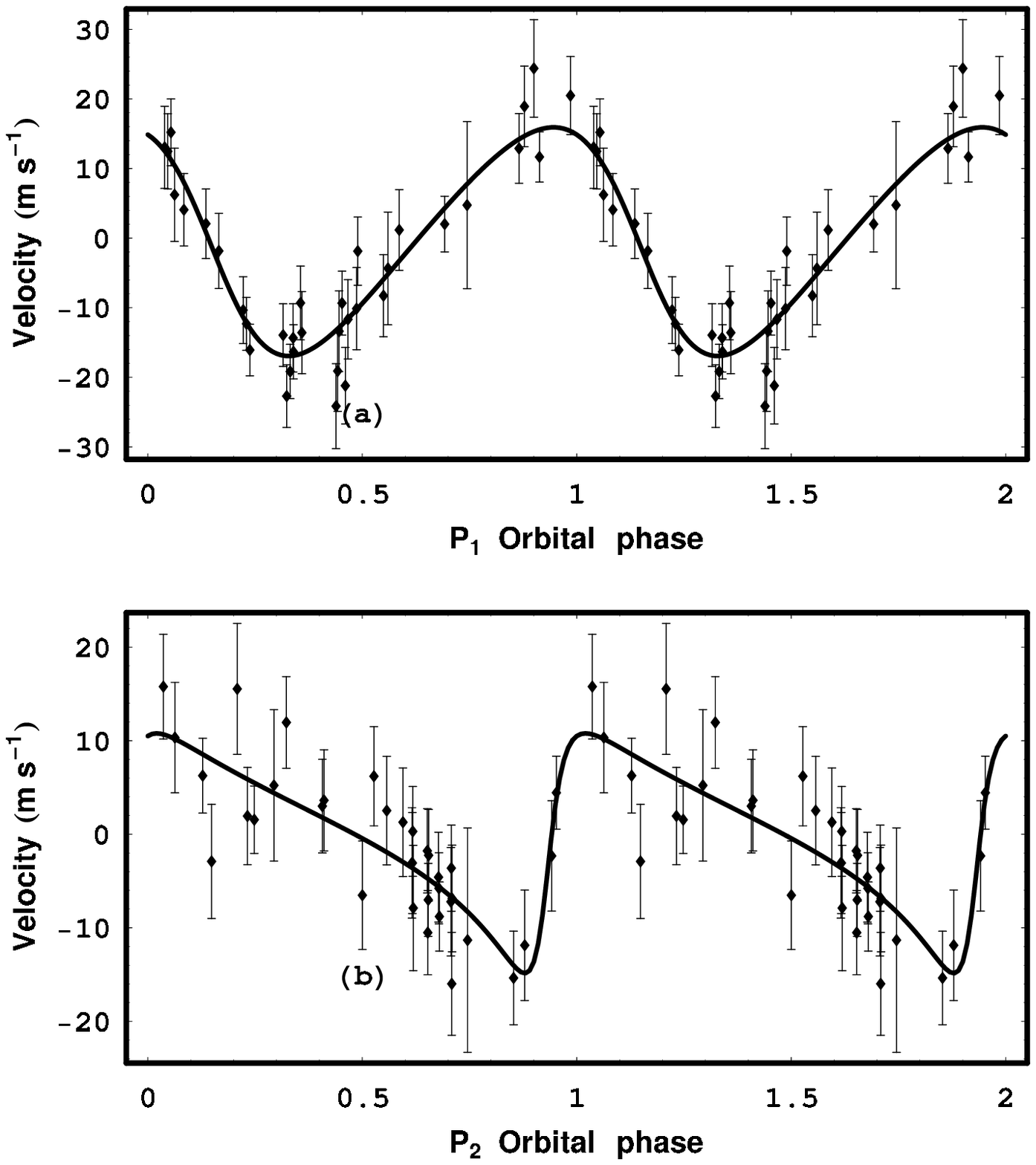}
 \caption{Panel (a) shows the data, with the best fitting $P_2$ orbit subtracted, for two cycles of $P_1$ phase with the best fitting $P_1$ orbit overlaid. Panel (b) shows the data plotted versus $P_2$ phase with the best fitting $P_1$ orbit removed.}
\label{fig:phase}
\end{figure}

Table~\ref{tab:parerrorsM2} gives our Bayesian two planet orbital parameter values and their errors for the two different choices of $s_0$. Apart from the $s$ parameter, changing $s_0$ from 1 to 10m s$^{-1}$ did not significantly alter the $M_2$ parameter estimates. We note that the MAP value for $P_2$ falls just outside the 68.3\% credible region. Finally, Panel (a) of Figure~\ref{fig:phase} shows the data, minus the best fitting $P_2$ orbit, for two cycles of $P_1$ phase. The best fitting $P_1$ orbit is overlaid. Panel (b) shows the data plotted versus $P_2$ phase with the best fitting $P_1$ orbit removed. The reduced $\chi^2 = 1.01$ for the best $M_2$ model fit when no jitter is assumed. The reduced $\chi^2 = 2.33$ for the best $M_1$ model fit with no jitter, and $1.13$ when a jitter of $5.4$m s$^{-1}$ is assumed. 

\begin{table*}
 \centering
 \begin{minipage}{140mm}
  \caption{One planet model parameter estimates.}
  \label{tab:parerrorsM1}
  \begin{tabular}{@{}lllllll@{}}
  \hline
   Parameter    &   Tinney et al. & Butler et al.& & Model $M_{1_j}$ & Model $M_1$ &  Model $M_1$  \\
   &    (2005) & (2006)  &  & & ($s_0 = 1$m s$^{-1}$) & ($s_0 = 10$m s$^{-1}$)\\
\hline
$P$  (d) & $130\pm1$ & $130.08\pm0.51$ & & $129.97_{-0.41}^{+0.45}$& $129.98_{-0.41}^{+0.42}$& $129.98_{-0.39}^{+0.45}$\\
& & & &(129.97)&(130.08)& (130.08)\\
& & & & & & \\
$K$ (m s$^{-1}$) & $20\pm2$ & $19.7\pm3.6$ & & $19.3_{-2.0}^{+1.9}$& $19.3_{-2.0}^{+2.0}$& $19.2_{-1.9}^{+2.1}$\\
& & & & (19.6)& (19.6)& (19.4) \\
& & & & & & \\
$e$ & $0.32\pm0.10$ & $0.24\pm0.16$ & & $0.22_{-.10}^{+.10}$ & $0.22_{-.11}^{+.10}$& $0.22_{-.11}^{+.11}$ \\
& & & & (0.24) & (0.24) & (0.23)\\
& & & & & & \\
$\omega$  (deg) & $126\pm40$ & $113$ & & $113_{-33}^{+32}$ & $114_{-31}^{+31}$&  $114_{-33}^{+31}$ \\
& & & & (116)& (111)& (113)\\
& & & & & & \\
$a$  (AU) & $0.49\pm0.04$ & $0.52\pm0.03$ & & $0.51_{-.02}^{+.02}$ & $0.51_{-.02}^{+.02}$ & $0.51_{-.02}^{+.02}
$ \\
& & & & (0.51) & (0.51)& (0.51) \\
& & & & & & \\
$M \sin i$  ($M_J$) & $0.45\pm0.05$ & $0.52\pm0.08$ & & $0.48_{-.05}^{+.05}$& $0.48_{-.06}^{+.06}$& $0.48_{-.06}^{+.06}$ \\
& & & & (0.49) & (0.49)& (0.49)\\
& & & & & & \\
Periastron & $11002.8\pm10$ & $10999\pm15$ & & $11001_{-10}^{+10}$ & $11000_{-11}^{+10}$& $11001_{-11}^{+10}$ \\
\ passage & & & & (10001) & (10999)& (10999)\\
\ (JD - 2,440,000) & & & & & & \\
& & & & &\\
$s$ (m s$^{-1}$) &  & & & $5.9_{-1.4}^{+1.2}$ & $5.6_{-1.5}^{+1.4}$ & $5.8_{-1.5}^{+1.4}$  \\
& & & & (4.5) & (4.6) & (4.4) \\
& & & & & & \\
RMS residuals & 7.2 & 8.2& & 7.5 & 7.5 & 7.5 \\
\ \  (m s$^{-1}$)& & & & & & \\
\hline
\end{tabular}
\end{minipage}
\end{table*}

\begin{table*}
 \centering
 \begin{minipage}{140mm}
  \caption{Two planet model parameter estimates.}
  \label{tab:parerrorsM2}
  \begin{tabular}{@{}llllll@{}}
  \hline
   Parameter    &  \multicolumn{2}{l}{Model $M_2$ ($s_0 = 1$m s$^{-1}$)} & &\multicolumn{2}{l}{Model $M_2$ ($s_0 = 10$m s$^{-1}$)} \\
   &   planet 1 & planet 2 & & planet 1 & planet 2\\
\hline
$P$  (d) & $129.8_{-0.4}^{+0.4}$ & $908_{-94}^{+81}$& & $129.8_{-0.4}^{+0.4}$ & $909_{-92}^{+82}$ \\
& (129.9)& (1001)& & (129.9)& (1001)\\
& & & & & \\
$K$ (m s$^{-1}$) & $16.5_{-1.6}^{+1.5}$ & $10.7_{-3.0}^{+2.1}$ &  & $16.5_{-1.5}^{+1.6}$ & $10.1_{-2.9}^{+2.2}$  \\
& (16.4) & (12.8) & & (16.4) & (12.8) \\
& & & & & \\
$e$ & $0.21_{-.09}^{+.09}$ & $0.38_{-.20}^{+.26}$ & & $0.20_{-.09}^{+.09}$ & $0.37_{-.20}^{+.26}$  \\
& (0.19) & (0.61) & & (0.19) & (0.61) \\
& & & & \\
$\omega$  (deg) & $123_{-37}^{+40}$ & $227_{-48}^{+53}$ & & $121_{-38}^{+49}$ & $226_{-52}^{+52}$ \\
& (99.7) & (255) & & (99.5) & (255)\\
& & & & & \\
$a$  (AU) & $0.51_{-.02}^{+.02}$ & $1.87_{-.15}^{+.13}$ & & $0.51_{-.02}^{+.02}$ & $1.87_{-.14}^{+.13}$  \\
& (0.510) & (1.991) & & (0.510) & (1.991) \\
& & & & \\
$M \sin i$  ($M_J$) & $0.413_{-.052}^{+.049}$ & $0.45_{-.11}^{+.09}$ & & $0.41_{-.05}^{+.05}$ & $0.45_{-.13}^{+.11}$ \\
& (0.414) & (0.513) & & (0.414) & (0.513)\\
& & & & & \\
Periastron & $11008_{-13}^{+13}$ & $10605_{-120}^{+148}$ & & $11007_{-13}^{+14}$ & $10601_{-125}^{+152}$ \\
\ passage &  (10995) & (10512)&  & (10998) & (10524)\\
\ (JD - 2,440,000) & & & & & \\
& & &&  & \\
$s$ (m s$^{-1}$) & $(0.0)_{-0.0}^{+1.6}$ & & & $(0.0)_{-0.0}^{+2.2}$ & \\
& & & & & \\
RMS residuals &  4.4 & & & 4.4 & \\
\ \  (m s$^{-1}$)& & & & & \\
\hline
\end{tabular}
\end{minipage}
\end{table*}

\section{Model selection}
\label{sec:modsel}

To compare the posterior probabilities of the two planet model to the one planet models we need to evaluate the odds ratio,
$O_{21} =p(M_{2} | D,I)/p(M_{1} | D,I)$, the ratio of the posterior probability
of model $M_{2}$ to model $M_{1}$.  Application of Bayes's 
theorem leads to,
\begin{equation}
O_{21} = {p(M_{2} | I) \over p(M_{1} | I)}\;
      {p(D | M_{2},I) \over p(D | M_{1},I)}
       \equiv {p(M_{2} | I) \over p(M_{1} | I)}\; B_{21}
\label{eq:orbit22}
\end{equation}
where the first factor is the prior odds ratio, and the second factor
is called the {\it Bayes factor}. The Bayes factor is the ratio of
the marginal (global) likelihoods of the models. The MCMC algorithm produces
samples which are in proportion to the posterior probability distribution which is fine for parameter
estimation but one needs the proportionality constant for estimating the model marginal likelihood. 
\citet{Clyde2006} recently reviewed the state of techniques for model selection from a statistics perspective and \citet{FordGregory2006} have evaluated the performance of a variety of marginal likelihood estimators in the extrasolar planet context. 

In this work we will compare the results from three marginal likelihood estimators: (a) parallel tempering, (b) ratio estimator, and (c) restricted Monte Carlo. The analysis presented in Section~\ref{sec:partempML}, \ref{sec:reML}, and \ref{sec:RMC} is based on the old data set (Tinney et al. 2005) and prior \#1. These results are summarized in Section~\ref{sec:summML} together with model selection results for the new data set (Butler et al. 2006) using prior \#2.

\subsection{Parallel tempering estimator}
\label{sec:partempML}

\begin{figure}
\includegraphics[width=80mm]{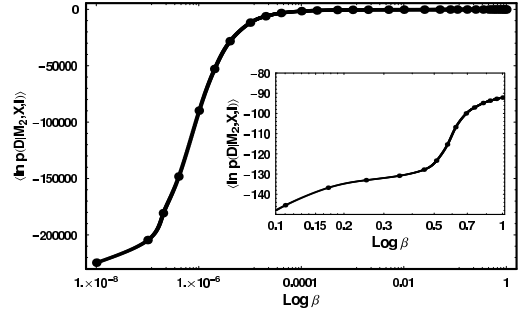}
 \caption{A plot of $\langle \ln[p(D|M_{i},X,I)]\rangle_{\beta}$ versus $\beta$. The inset shows a blow-up of the range $\beta= 0.1$ to 1.0.}
\label{fig:bayesfactor}
\end{figure}

The MCMC samples from all ($n_\beta$) simulations can be used to calculate the marginal likelihood of a model
according to equation~(\ref{eq:Zbeta10}) \cite{Gregorybook}.
\begin{equation}
\ln[p(D|M_{i},I)] =  \int d\beta \langle \ln[p(D|M_{i},X,I)]\rangle_{\beta},
\label{eq:Zbeta10}
\end{equation}
where $i = 0,1,2$ corresponds to a zero, one or two planet model, and $X$ represent a vector of the model parameters which includes the extra Gaussian noise parameter $s$. In words, for each of the $n_\beta$ parallel simulations, compute the expectation value (average) of the natural logarithm of the likelihood for post burn-in MCMC samples. 
It is necessary to use a sufficient number of tempering levels that we can estimate the above integral by interpolating values of 
\begin{equation}
\langle \ln[p(D|M_{i},X,I)]\rangle_{\beta}= \frac{1}{n}\sum_t \ln[p(D|M_{i},X,I)]_{\beta},
\label{eq:Zbeta10a}
\end{equation}
in the interval from $\beta = 0$ to 1, from the finite set. For this problem we used 34 tempering in the range $\beta = 10^{-8}$ to 1.0. Figure~\ref{fig:bayesfactor} shows a plot of $\langle \ln[p(D|M_{i},X,I)]\rangle_{\beta}$ versus $\beta$. The inset shows a blow-up of the range $\beta= 0.1$ to 1.0. 

The relative importance of different decades of $\beta$ can be judged from Table~\ref{tab:FracError}.
The second column gives the fractional error that would result if this decade of $\beta$ was not included and thus indicates the sensitivity of the result to that decade. In \citet{FordGregory2006} we constructed a similar table for the one planet system HD 88133 only we investigated a wider range of $\beta$ values down to $10^{-11}$. The fractional errors for the $\beta$ range $10^{-5}$ to $10^{-8}$ were very similar to those given in Table~\ref{tab:FracError}. The fractional error for $\beta = 10^{-11}\ - \ 10^{-8}$ was only $0.002$ and consequently this range of $\beta$ can safely be ignored.
\begin{table}
 \centering
 \begin{minipage}{160mm}
  \caption{Fractional error versus $\beta$ for the results shown in Figure~\ref{fig:bayesfactor}.}
  \label{tab:FracError}
  \begin{tabular}{@{}lllll@{}}
  \hline
   $\beta$ range    &    Fractional error \\
\hline
$1.0\ -\ 10^{-1}$      & $1.55 \times 10^{44}$  \\
$10^{-1}\ -\ 10^{-2}$  & $7.46 \times 10^{6}$  \\
$10^{-2}\ -\ 10^{-3}$  & $15.2$  \\
$10^{-3}\ -\ 10^{-4}$  & $1.17$  \\
$10^{-4}\ -\ 10^{-5}$  & $0.52$  \\
$10^{-5}\ -\ 10^{-6}$  & $0.35$  \\
$10^{-6}\ -\ 10^{-7}$  & $0.16$  \\
$10^{-7}\ -\ 10^{-8}$  & $0.02$  \\
\hline
\end{tabular}
\end{minipage}
\end{table}

\begin{figure}
\includegraphics[width=80mm]{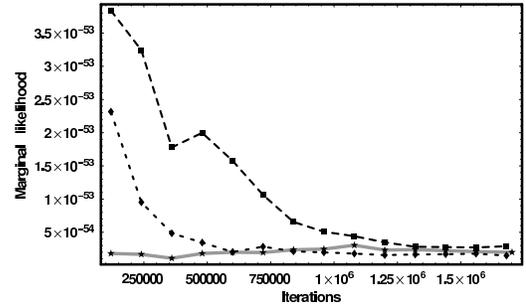}
 \caption{A plot of the parallel tempering marginal likelihood versus iteration number (dashed curves) for two different MCMC runs for the two planet model. The solid gray curve shows the result for the ratio estimator method which is discussed in Section~\ref{sec:reML}.}
\label{fig:bayesfactor1}
\end{figure}

Figure~\ref{fig:bayesfactor1} shows a plot of the parallel tempering marginal likelihood versus iteration number for two different MCMC runs for the two planet model. The solid gray curve shows the result for the second ratio estimator method which is discussed below. All three curves converge to the same value but the ratio estimator converges much more rapidly. The final parallel tempering marginal likelihood estimates from the two runs were $1.5 \times 10^{-54}$ and $3.4 \times 10^{-54}$, yielding an average value of $2.4 \times 10^{-54}$.

\subsection{Marginal likelihood ratio estimator}
\label{sec:reML}

Our second method was introduced by \citet{FordGregory2006}. It makes use of an additional sampling distribution $h(X)$. Our starting point is Bayes' theorem
\begin{equation}
p(X|M_{i},I) = \frac{p(X|M_i,I) p(D|M_i,X,I)}{p(D|M_i,I)}.
\label{eq:marglike1}
\end{equation}
Re-arranging the terms and multiplying both sides by $h(X)$ we obtain
\begin{eqnarray}
& & p(D|M_i,I) p(X|M_{i},I) h(X) = \nonumber \\
& & \ \ \ \ \ \ \ \ \ \ \ p(X|M_i,I) p(D|M_I,X,I) h(X).
\label{eq:marglike2}
\end{eqnarray}
Integrate both sides over the prior range for $X$.
\begin{eqnarray}
& & p(D|M_i,I)_{re} \int p(X|M_{i},I) h(X) dX =  \nonumber \\
& & \ \ \ \ \ \ \ \ \ \ \int p(X|M_i,I) p(D|M_I,X,I) h(X) dX.
\label{eq:marglike3}
\end{eqnarray}
The ratio estimator of the marginal likelihood, which we designate by $p(D|M_i,I)_{re}$, is given by 
\begin{equation}
p(D|M_{i},I)_{re} = \frac{\int p(X|M_i,I) p(D|M_i,X,I) h(X) dX}{\int p(X|M_{i},I) h(X) dX}.
\label{eq:marglikeRE}
\end{equation}
To obtain the marginal likelihood ratio estimator, $p(D|M_{i},I)_{re}$, we approximate the numerator by drawing samples $\tilde{X}^1,\tilde{X}^2,\cdots,\tilde{X}^{n_s^{\prime}}$ from $h(X)$ and approximate the denominator by drawing samples $\vec{X}^1,\vec{X}^2,\cdots,\vec{X}^{n_s}$ from the $\beta = 1$ MCMC post burn-in iterations. The arbitrary function $h(X)$ was set equal to a multinormal with a covariance matrix equal to twice the covariance matrix computed from a sample of the $\beta=1$ MCMC output. We used~\footnote{According to \citet{FordGregory2006}, the numerator converges more rapidly than the denominator.} $n_s^\prime = 10^5$ and $n_s$ from $10^4$ to $2 \times 10^5$. Some of the samples from a multinormal $h(X)$ can have nonphysical parameter values (e.g. $K <0$). Rejecting all nonphysical samples corresponds to sampling from a truncated multinormal. The factor required to normalize the truncated multinormal is just the ratio of the total number of samples from the full multinormal to the number of physical valid samples. Of course we need to use the same truncated multinormal in the denominator of equation~(\ref{eq:marglikeRE}) so the normalization factor cancels. From Figures~\ref{fig:bayesfactor} and \ref{fig:bayesfactor1} it is clear that the $p(D|M_{2},I)_{re}$ converges much more rapidly than the parallel tempering estimator and the parallel tempering estimator, $p(D|M_{2},I)_{PT}$, required 34 $\beta$ simulations instead of one. The final values of $p(D|M_{i},I)_{re}$ for models $M_{2}$ and $M_{1}$ were $2.0 \times 10^{-54}$ and $3.3 \times 10^{-56}$. The solid gray curve in Figure~\ref{fig:bayesfactor1} shows marginal likelihood ratio estimator versus iteration number for model $M_2$. It lies between the two parallel tempering convergence values.

\subsection{Restricted Monte Carlo marginal likelihood estimate}
\label{sec:RMC}

We can also make use of Monte Carlo integration to evaluate the marginal likelihood as given by equation~(\ref{eq:marglike4}).
\begin{equation}
p(D|M_i,I) = \int p(X|M_i,I) p(D|M_I,X,I) dX.
\label{eq:marglike4}
\end{equation}
Monte Carlo (MC) integration can be very inefficient in exploring the whole prior parameter range, but once we have established the significant regions of parameter space with the MCMC results, this is no longer the case.
The outer borders of the MCMC marginal parameter distributions were used to delineate the boundaries of the volume of parameter space to be used in the Monte Carlo integration. RMC integration using $10^7$ samples was repeated three times for the two planet model. The results were $1.6\pm 2.6, 1.1\pm 1.2, 2.3\pm 0.6 \times 10^{-54}$ yielding a weighted average of $2.1\pm 0.1 \times 10^{-54}$. The weighted average of RMC repeats for the one planet model was $2.9\pm 0.1 \times 10^{-56}$.

\subsection{Summary of model selection results}
\label{sec:summML}

Table~\ref{tab:modelSelold} summarizes the marginal likelihoods and Bayes factors comparing models $M_{0}$ and $M_{2}$ to $M_{1}$ for the old data set analyzed using prior \#1. For model $M_{0}$, the marginal likelihood was obtained by numerical integration.  For $M_{1}$, the value and error estimate are based on the ratio estimator and RMC methods. For model $M_2$, the two parallel tempering marginal likelihood estimates differed by a factor of $\sim 2$, although their average agreed within $20\%$ with the values obtained by the ratio estimator and RMC methods. Combining the three results yielded a marginal likelihood good to $\sim \pm 10\%$. The conclusion is that the Bayes factor strongly favors a two planet model compared to the other two models. 

Table~\ref{tab:modelSelnew} gives the same information for the new data set analyzed using the improved prior \#2 for both choices of $s_0$ ($1$ and $10$m s$^{-1}$). The improved data significantly strengthens the case for the existence of the second planet as judged by the value of $B_{21}$. Since the value of $B_{21}$ decreases as we increase the value of $s_0$, what argument can we give for not using a very much larger value, say $s_0 = 1000$. This would effectively convert our Jeffreys prior into a uniform prior. For a uniform prior, the probability in the decade $s = 100$ to 1000m s$^{-1}$ is 100 times the probability in the decade from 1 to 10m s$^{-1}$. Thus, a uniform prior assumes that it is much more likely that $s$ is very large than it is very small. We argue that the scale invariant property of the Jeffreys prior is much more consistent with our prior state of knowledge for the current problem. For model selection purposes we want to employ the smallest value of $s_0$ that avoids the divergence that would otherwise occur at $s=0$. For the current problem, a value of $s_0=10$m s$^{-1}$ is large enough to achieve this goal.

\begin{table}
 \centering
  \caption{Marginal likelihoods for old data set (Tinney et al. 2005) and prior \#1.}
  \label{tab:modelSelold}
  \begin{tabular}{@{}llll@{}}
  \hline
  Model & $s_0$ & Marginal & Bayes  \\
         & (m s$^{-1}$)& Likelihood & factor \\
\hline
$M_{0}$ & 1.0  & $1.60 \times 10^{-60}$ &  $(76\pm 7) \times 10^{-6}$ \\
$M_{1}$ & 1.0 & $(3.10\pm 0.3)\times 10^{-56}$ &  $1.0$ \\
$M_{2}$ & 1.0 & $(2.2\pm 0.2)\times 10^{-54}$ &  $(71\pm 9)$ \\
\hline
\end{tabular}
\end{table}
\begin{table}
 \centering
  \caption{Marginal likelihoods for new data set (Butler et al. 2006) and prior \#2.}
  \label{tab:modelSelnew}
  \begin{tabular}{@{}llll@{}}
  \hline
   Model & $s_0$ & Marginal & Bayes  \\
         & (m s$^{-1}$)& Likelihood & factor \\
\hline
$M_{0}$ & 1.0 & $1.56 \times 10^{-68}$ &  $(38\pm 4) \times 10^{-6}$ \\
$M_{1}$ & 1.0 & $(4.08\pm 0.39)\times 10^{-64}$ &  $1.0$ \\
$M_{2}$ & 1.0 & $(9.54\pm 0.87)\times 10^{-62}$ &  $234\pm 31$ \\
& & &\\
$M_{1_j}$ &  & $(2.44\pm 0.24)\times 10^{-63}$ &  \\
& & &\\
$M_{0}$ & 10.0 & $1.44 \times 10^{-68}$ &  $(58\pm 6) \times 10^{-6}$ \\
$M_{1}$ & 10.0 & $(2.47\pm 0.25)\times 10^{-64}$ &  $1.0$ \\
$M_{2}$ & 10.0 & $(4.38\pm 0.69)\times 10^{-62}$ &  $177\pm 33$ \\
\hline
\end{tabular}
\end{table}

Table~\ref{tab:modelSelnew} gives values of $B_{21}$ and $B_{01}$ for the two cases, $s_0=1$ and 10m s$^{-1}$. Table~\ref{tab:modelSelnew} also includes an entry for $M_{1_j}$, the one planet plus stellar jitter model (introduced at the end of section~\ref{sec:priors}). It is not as simple to compare models $M_{1_j}$ to $M_2$ because $M_{1_j}$ makes use of additional prior information concerning stellar velocity noise or jitter. One approximate way to make use of this information for model $M_2$ would be to set the prior upper bound on our extra noise parameter at a few times the jitter estimate, say 15m s$^{-1}$. Reducing the prior upper bound on $s$ from 2129m s$^{-1}$ to 15m s$^{-1}$ reduces the Occam penalty associated with the unknown $s$ parameter by a factor of $\approx 6$ for $s_0=10$m s$^{-1}$. Thus for $s_0=10$ m s$^{-1}$, the marginal likelihood for $M_2$ would need to be increased by a factor of $\approx 6$ before comparing with the likelihood for $M_{1_j}$. The corresponding Bayes factor is $B_{21_j} \approx 108$. 

Following \citet{Cumming2004}, we compute a Bayesian false alarm probability. For this purpose we will restrict the hypothesis space of interest to just two models $M_{1}$ and $M_2$. In this context, the Bayesian false alarm probability, $F$, is the probability there is really only one planet given the data $D$ and our prior information $I$. In this case the probability of the data is 
\begin{equation}
p(D|I)=p(M_{1}|I)p(D|M_{1},I)+p(M_2|I)p(D|M_2,I). 
\label{eq:pd}
\end{equation}
Combine this with Bayes theorem
\begin{equation}
p(M_{1}|D,I)=p(M_{1}|I)p(D|M_{1},I)/p(D|I), 
\label{eq:pd1}
\end{equation}
to obtain
\begin{equation}
F=p(M_{1}|D,I)=\frac{1}{1+\frac{p(M_2|I)}{p(M_{1}|I)}\frac{p(D|M_2,I)}{p(D|M_{1},I)}}= \frac{1}{1+B_{21}},
\label{eq:pd2}
\end{equation}
where we have assumed $p(M_{1}|I)=p(M_2|I)$. 
According to equation~(\ref{eq:pd2}), a Bayes factor $B_{21} \approx 177$ corresponds to a false alarm probability of 0.006.

\section{Discussion}

One source of error in the measured velocities is "jitter", which is due in part to flows and inhomogeneities on the stellar surface. Wright (2005) gives a model that estimates, to within a factor of roughly 2 \citep{Butler2006}, the jitter for a star based upon a star's activity, color, Teff, and height above the main sequence. For HD 208487, \citet{Butler2006} quote a jitter estimate of $5.4$m s$^{-1}$, based on Wright's model. Our models $M_0$, $M_1$ and $M_2$ employ instead an extra Gaussian noise nuisance parameter, $s$, with a prior upper bound of equal to $K_{\rm max}=2129$m s$^{-1}$. Anything that cannot be explained by the model and published measurement uncertainties (which do not include jitter) contributes to the extra noise term. Of course, if we are interested in what the data have to say about the size of the extra noise term then we can readily compute the marginal posterior for $s$. The marginal for $s$ is shown in Figures~\ref{fig:marg2} and \ref{fig:marg1} for models $M_2$ and $M_1$, respectively. For $M_1$, the marginal for $s$ shows a pronounced peak with a median value of $5.6$m s$^{-1}$, which is very close to the jitter estimate given in \citet{Butler2006} based on Wright's model. 

\begin{figure}
\includegraphics[width=80mm]{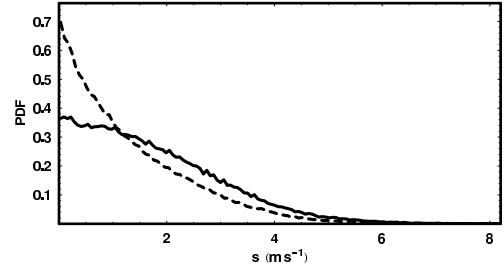}
 \caption{A comparison of the marginal probability distributions for the extra noise parameter, $s$, for
two different values of the prior knee, $s_0$ for model $M_2$. The dashed curve corresponds to $s_0 = 1.0$m s$^{-1}$ and the solid curve to $s_0 = 10.0$m s$^{-1}$.}
\label{fig:s0knee}
\end{figure}

For $M_2$, the marginal for $s$ shows a sharp peak at a value of $s=0$m s$^{-1}$. We can determine whether the posterior shape was strongly influenced by our choice of knee ($s_0 = 1.0$m s$^{-1}$) in the modified Jeffreys prior for $s$, by comparing with the results obtained assuming $s_0 = 10$m s$^{-1}$. The two marginal posteriors are shown in Figure~\ref{fig:s0knee}. Clearly, using a larger $s_0$ does suppress the sharp peak at $s = 0$, however, the marginal posterior still favors a value of $s$ close to zero. It would appear that the velocity variations arising from the second planet are on the same scale as the previously estimated stellar jitter for HD 208487. The results of our Bayesian model selection analysis indicate that a two planet model is greater than 100 times more probable than a one planet model with the previously estimated jitter. Based on model $M_2$, and $s_0 = 10$m s$^{-1}$, the $95\%$ upper limit on the remaining stellar jitter is $4.2$m s$^{-1}$. 

\begin{figure}
\includegraphics[width=80mm]{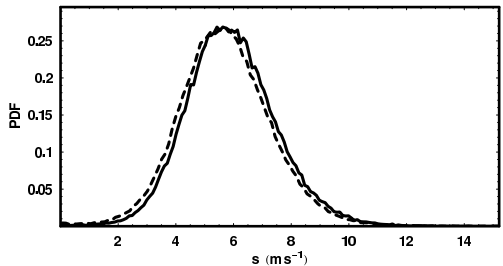}
 \caption{A comparison of the marginal probability distributions for the extra noise parameter, $s$, for
two different values of the prior knee, $s_0$ for model $M_1$. The dashed curve corresponds to $s_0 = 1.0$m s$^{-1}$ and the solid curve to $s_0 = 10.0$m s$^{-1}$.}
\label{fig:s0knee1}
\end{figure}

Figure~\ref{fig:s0knee1} shows a comparison of the marginal for $s$ for $s_0=1$m s$^{-1}$ (dashed curve) and $s_0=10$m s$^{-1}$ (solid) for model $M_1$. In this case, the marginal for $s_0=10$m s$^{-1}$ is displaced very slightly to larger values of $s$, as expected for a more uniform prior, but the shift is negligible when compared to the uncertainty in the parameter value. 

It is interesting to compare the measured one $\sigma$ width of the marginal period PDF to values estimated from equation~(\ref{eq:Bfres}) after multiplying by $P^2$ to convert to $\delta f$ to $\delta P$. We have used the ratio of $K/\sqrt{2}$ to the mean velocity error as an estimate of the $S/N$ for use in equation~(\ref{eq:Bfres}). For $P_1$ the predicted $\delta P_{\rm pred} = 0.31$, whereas the measured value $\delta P_{\rm meas} = 0.4$. Of course, the effect of any correlations between parameters is to broaden the marginal distribution, so if anything we expect the measured value to be broader. In the case of $P_2$, $\delta P_{\rm meas} \approx 1.8 \times \delta P_{\rm pred}$. In this case both the shape of the marginal and the $P_2$ versus $e_2$ correlation diagram of Figure~\ref{fig:corr} suggest that the data allow for two closely blended solutions in the vicinity of 900d. The marginals for $\chi_2$ and $\omega_2$ also show evidence for a blend of two components. A preliminary three planet model analysis failed to detect a third planet.

A rough estimate of the expected width of the marginal velocity PDF is equal to the mean velocity error divided by the $\sqrt{N}$, or $0.9$m s$^{-1}$, where $N$ is the number of data points. In comparison the measured widths are 1.6m s$^{-1}$ for $P_1$ and 2.5m s$^{-1}$ for $P_2$. 

In section~\ref{sec:priors}, we discussed two different strategies to search the orbital frequency parameter space for a multi-planet model: (a) an upper bound on $f_1 \le f_2 \le \cdots \le f_n$  is utilized to maintain the identity of the frequencies, and (b) all $f_i$ are allowed to roam over the entire frequency range  and the parameters re-labeled afterwards. In this second case, nothing constrains $f_{i-1}$ to be less than $f_i$ so that degenerate parameter peaks can occur. For case (b) and an $n$ planet model, the number of possible peaks is $n!$ more than in (a). In this work we adopted approach (b) because in repeated blind frequency (period) searches it always converged on the highest posterior probability peak, in spite of the huge period search range. Approach (a) failed to locate the highest peak in some trials and in the cases where it succeeded it required many more iterations. 

Apart from their very different relative efficiency to locate the highest probability peak, the two approaches are equivalent and equation~(\ref{eq:freqs4}) is the appropriate frequency prior for both approaches. As a test of this claim, we obtained $p(D|M_2,I)_{re} = 1.9 \times 10^{-54}$ from an MCMC run using approach (a), the old data set, and prior \#1. This compares closely with the value $2.0 \times 10^{-54}$ obtained for approach (b).
Again, from an MCMC run using prior \#2 and the old data set, we obtained $p(D|M_2,I)_{re} = 3.9 \times 10^{-54}$ for (a) and a value of $p(D|M_2,I)_{re} = 3.9 \times 10^{-54}$ for (b). Further tests of this claim were carried out in a entirely different problem involving the detection of three spectral lines. For this problem the marginal likelihoods computed using approaches (a) and (b) agreed to within 1\%.

It is interesting to compare the performance of the three marginal likelihood estimators employed in this work to their performance on another data set for HD 88133, which involved fitting a one planet model \citep{FordGregory2006}. In the latter study, 5 separate ($n\beta = 32$) parallel tempering runs were carried out. After 300,000 iterations, four estimates agreed within $25\%$ while the fifth was a factor of two larger. Based on these two applications, it would appear that any single parallel tempering marginal likelihood estimate is only accurate to a factor of 2. For HD 208487 and a two planet model, the parallel tempering estimator required $\sim 1.5 \times 10^6$ iterations for convergence. Of course to establish convergence, it is necessary to execute at least two separate parallel tempering runs. The chief advantage of parallel tempering, in respect to model selection, is that it is able to estimate the marginal likelihood even for multimodal posterior distributions. For HD 88133, both the marginal likelihood ratio estimator and RMC estimator yielded values within $5\%$ of the average of the best 4 parallel tempering estimates. For the 2 planet model of HD 208487 and only two parallel tempering runs, the three estimators agreed within $20\%$, with the final estimate considered accurate to $\pm 10\%$.  For more information on these and additional marginal likelihood estimators see \citet{FordGregory2006}. 

\section{Conclusions}

In this paper we demonstrate the capabilities of an automated Bayesian parallel tempering MCMC approach to the analysis of precision radial velocities. The method is called a Bayesian Kepler periodogram because it is ideally suited for detecting signals that are consistent with Kepler's laws. However, it is more than a periodogram because it also provides full marginal posterior distributions for all the orbital parameters that can be extracted from radial velocity data. The periodogram employs an alternative method for converting the time of an observation to true anomaly that enables it to handle much larger data sets without a significant increase in computation time. 

A preliminary re-analysis of the old data for HD 208487 found evidence for a second planet \citep{Gregory2005MaxEnt}. This has now been confirmed by the current analysis based on the results of the improved Kepler periodogram and the new data set of Butler et al. (2006). The velocity variations arising from the second planet are on the same scale as the previously estimated stellar jitter for HD 208487. Based on our two planet model results, the $95\%$ upper limit on stellar jitter is $4.2$m s$^{-1}$.

We also derived the form of joint frequency prior to use for a multiple planet search which is given by
\begin{equation}
p(f_1,f_2, \cdots f_n|M_n,I) = \frac{n!}{f_1 f_2 \cdots f_n [\ln (f_H/f_L)]^n} 
\label{eq:freqs5n}
\end{equation} 
where $n$ is the number of planets. There are two frequency search strategies: (a) constrain the frequencies such that $f_L < f_1 < f_2 < \cdots < f_n < f_H$, or (b) allow each frequency to be unconstrained and re-label them afterwards. In practice we found that case (b) was significantly more successful than (a) for blind searches. In both cases equation~(\ref{eq:freqs5n}) is the correct joint frequency prior.

Considerable attention was paid to the topic of Bayesian model selection. Based on current research, it appears necessary to employ more than one method for estimating marginal likelihoods in order to obtain results that are accurate at the $10\%$ level.

\section*{Acknowledgments}

I thank the members of the exoplanet working group during the Astrostatistics program at the Statistical and Applied Mathematical Sciences Institute for many useful discussions. This research was supported in part by grants from the Canadian Natural Sciences and Engineering Research Council at the University of British Columbia.

\bsp

\label{lastpage}


\section{Bibliography}

\begin{thebibliography}{99}

\bibitem[\protect\citeauthoryear{Bretthorst}{1988}]{Brett1988} Bretthorst, G. L., 1988, Bayesian Spectrum Analysis and Parameter
Estimation, New York: Springer-Verlag

\bibitem[\protect\citeauthoryear{Bretthorst}{2001}]{Brett2001} Bretthorst, G.L., 2001, Ali Mohammad-Djafari, ed, AIP Conference Proceedings, New York, 568, 241

\bibitem[\protect\citeauthoryear{Bretthorst}{2003}]{Brett2003} Bretthorst, G. L., 2003, in `Bayesian Inference and Maximum Entropy Methods in Science and Engineering', C.W. Williams, ed, American Institute of Physics Conference Proceedings, 659, 3


\bibitem[\protect\citeauthoryear{Butler et al.}{2006}]{Butler2006} Butler, R. P., Wright, J. T., Marcy, G. W., Fischer, D. A., Vogt, S. S., Tinney, C. G., Jones, H. R. A., Carter, B. D., Johnson, J. A., McCarthy, C., and Penny, A. J., 2006, ApJ, 646, 505

\bibitem[\protect\citeauthoryear{Cumming}{2004}]{Cumming2004} Cumming, A. 2004, MNRAS, 354, 1165

\bibitem[\protect\citeauthoryear{Clyde}{2006}]{Clyde2006} Clyde, M., in `Statistical Challenges in Modern Astronomy IV,'  G. J. Babu and E. D. Feigelson (eds.), San Francisco:Astron. Soc. Pacific (in press 2006) 
 
 

\bibitem[\protect\citeauthoryear{Ford}{2005}]{Ford2005} Ford, E. B., 2005, AJ, 129, 1706

\bibitem[\protect\citeauthoryear{Ford}{2006}]{Ford2006} Ford, E. B., 2006, ApJ, 620, 481

\bibitem[\protect\citeauthoryear{Ford \& Gregory}{2006}]{FordGregory2006} Ford, E. B., \& Gregory, P. C., in `Statistical Challenges in Modern Astronomy IV',  G. J. Babu and E. D. Feigelson, eds, San Francisco:Astron. Soc. Pacific (in press 2006) astro-ph 0608328  
      

\bibitem[\protect\citeauthoryear{Gregory}{1999}]{Gregory1999} Gregory, P. C., 1999, ApJ, 520, 361 

\bibitem[\protect\citeauthoryear{Gregory}{2005a}]{Gregorybook} Gregory, P. C., 2005a, `Bayesian Logical Data Analysis for the Physical Sciences: A Comparative approach with {\it Mathematica} Support', Cambridge University Press

\bibitem[\protect\citeauthoryear{Gregory}{2005b}]{Gregory2005b} Gregory, P. C., 2005b, ApJ, 631, 1198

\bibitem[\protect\citeauthoryear{Gregory}{2005c}]{Gregory2005MaxEnt} Gregory, P. C.,2005c, in `Bayesian Inference and Maximum Entropy Methods in Science and Engineering', San Jose, eds. A. E. Abbas, R. D. Morris, J. P.Castle, AIP Conference Proceedings, 803, 139

\bibitem[\protect\citeauthoryear{Gregory \& Loredo}{1992}]{GregoryLoredo1992} Gregory, P. C., and Loredo, T. J., 1992, ApJ, 398, 146

\bibitem[\protect\citeauthoryear{Jaynes}{1957}]{Jaynes1957} Jaynes, E. T., 1957, Stanford University Microwave Laboratory Report 421, Reprinted in `Maximum Entropy and Bayesian Methods in Science and Engineering', G. J. Erickson and C. R. Smith, eds, (1988) Dordrecht: Kluwer Academic Press, p.1


\bibitem[\protect\citeauthoryear{Loredo}{2004}]{Loredo2004} Loredo, T., 2003, in `Bayesian Inference And Maximum Entropy Methods in Science and Engineering: 23rd International Workshop', G.J. Erickson \& Y. Zhai, eds, AIP Conf. Proc. 707, 330 (astro-ph/0409386).

\bibitem[\protect\citeauthoryear{Loredo \& Chernoff}{2003}]{LoredoChernoff2003} Loredo, T. L. and Chernoff, D., 2003, in `Statistical Challenges in Modern Astronomy III', E. D. Feigelson and G. J. Babu, eds , p. 57
  
\bibitem[\protect\citeauthoryear{Roberts et al.}{1997}]{Roberts1997} Roberts, G. O., Gelman, A. and Gilks, W. R., 1997, Annals of Applied Probability, 7, 110

\bibitem[\protect\citeauthoryear{Saar \& Donahue}{1997}]{Saar1997} Saar, S. H., \& Donahue, R. A., 1997, ApJ, 485, 319

\bibitem[\protect\citeauthoryear{Saar et al.}{1998}]{Saar1998} Saar, S. H., Butler, R. P., \& Marcy, G. W. 1998, ApJ, 498, L153

\bibitem[\protect\citeauthoryear{Scargle}{1982}]{Scargle1982} Scargle, J. D. 1982, ApJ, 263, 835

\bibitem[\protect\citeauthoryear{Tinney et al.}{2003}]{Tinney2003} Tinney, C. G., Butler, R. P., Marcy, G. W., Jones, H. R. A., Penny, A. J., McCarthy, C., Carter, B. D., \& Bond, J., 2003, ApJ, 587, 423

\bibitem[\protect\citeauthoryear{Tinney et al.}{2005}]{Tinney2005} Tinney, C. G., Butler, R. P., Marcy, G. W., Jones, H. R. A., Penny, A. J., McCarthy, C., Carter, B. D., \& Fischer, D. A., 2005, ApJ, 623, 1171

\bibitem[\protect\citeauthoryear{Valenti \& Fischer}{2005}]{Valenti2005} Valenti, J. A., \& Fischer, D. A. 2005, ApJS, 159, 141

\bibitem[\protect\citeauthoryear{Wright}{2005}]{Wright2005} Wright, J. T., 2005, PASP, 117, 657

\bibitem[\protect\citeauthoryear{Wolszan \& Frail}{1992}]{Wolszan1992} Wolszcan, A., \& Frail, D., 1992, Nature,355, 145

\end{thebibliography}
\end{document}